\documentstyle[12pt]{article}
\newcommand{\be}{\begin{equation}}
\newcommand{\ee}{\end{equation}}
\newcommand{\ba}{\begin{eqnarray}}
\newcommand{\ea}{\end{eqnarray}}

\date{}

\begin{document}
\begin{center}
{\Large \bf Classical Integrable 2-dim Models Inspired by
SUSY Quantum Mechanics.}\\
\vspace{0.5cm}
{\large \bf A.A.Andrianov\footnote{E-mail: andrian@snoopy.phys.spbu.ru},
M.V.Ioffe\footnote{E-mail: ioffe@snoopy.phys.spbu.ru} , D.N.Nishnianidze}\\
Sankt-Petersburg State University,198904 Sankt-Petersburg, Russia
\end{center}
\vspace{0.5cm}
\hspace*{0.5in}
\begin{minipage}{5.0in}
{\small
A class of integrable 2-dim classical systems with integrals of
motion of fourth order in momenta is obtained from the quantum
analogues with the help of deformed SUSY algebra.
With similar technique a new class of potentials connected with
Lax method is found which provides the integrability of
corresponding 2-dim hamiltonian systems. In addition, some integrable
2-dim systems with potentials expressed in elliptic functions are
explored.}
\end{minipage}

\section{\bf Introduction}
\vspace*{0.5cm}
Construction of classical integrable systems with additional
integrals of motion is of considerable interest in Mathematical
Physics (see \cite{1} and references therein). Multidimensional
integrable systems play an important role to describe the dynamics 
 analogously to 1-dim
manifestly integrable systems. In particular, they may serve as
zero-approximations of perturbation theory in the case of weak,
nonintegrable perturbations.
Variety of
traditional approaches to this problem exists starting from
J.Kepler, S.Kowalewski till the Lax method. On the other hand
a modern
viewpoint on how to build classical integrable systems is based
on the symmetries of related quantum systems \cite{2}.
Recently the method for searching quantum integrable 2-dim systems
was developed \cite{3},\cite{9} with the help of a deformed
supersymmetry (SUSY) algebra formed by intertwining differential
operators of finite order.
Supersymmetry \cite{4}-\cite{6}, i.e.
the construction of the isospectral pair of Hamiltonians, was proved
\cite{3} to be in one-to-one correspondence to integrability of both
Hamiltonians, i.e. to existence of a differential symmetry operator,
which is polynomial in derivatives and which transforms solutions of
2-dim Schr\"odinger equation into other solutions with the same energy.
Quasiclassical reduction of the deformed SUSY algebra \cite{3} gave
the factorization of classical integrals of motion for the corresponding
Hamiltonians \cite{7}. As a result, the structure of analytically
resolved integrals of motion became more clear, and new classes of
integrable potentials were found \cite{7},\cite{8}.

In our paper we continue our study \cite{8} of classical systems which
integrability is
induced or inspired by a deformed SUSY algebra for the relevant quantum
systems. The consise basic construction of systems possessing a dynamical
symmetry  with the help of Higher derivative SUSY algebra is essentially
supplemented with algorithms of searching analytical solutions of related
nonlinear equations for coefficients functions of symmetry operators and
potentials.

In Sect.2 the intertwining relations between a pair of quantum
Schr\"odinger-type Hamiltonians by general differential
operators $ q^{\pm} $ of second order are investigated. The class
of particular solutions of these relations is constructed for the
cases of hyperbolic (Lorentz) $ g_{ik}=diag(1,-1) $ and degenerate
$ g_{ik}=(1,0) $ metric structures of operators $ q^{\pm} $ in second
derivatives. The differential operators of fourth order in derivatives,
which are symmetry operators for intertwined Hamiltonians, are
built. In Sect.3 the classical limit $ \hbar\to 0 $ for the
Hamiltonians is considered, and the class of systems with
integrals of motion of fourth order in momenta is obtained.
In Section 4 a new class of integrable systems with potentials
connected to the Lax method is derived using ansatzes
and technique taken from 2-dim SUSY Quantum Mechanics (Section 2).
Last Section 5 is devoted to description of some integrable systems
expressed in elliptic functions. We stress that quite a few of the
obtained potentials do not allow the separation of variables in known
coordinate systems and some of them sofar have not been found.

\section{\bf Quantum integrable 2-dim systems.}

In the 2-dimensional generalization \cite{3}, \cite{7}, \cite{9}
of Higher Order SUSY Quantum Mechanics \cite{10} the intertwining
relations of second order in derivatives are most essential:
\ba
H^{(1)}q^+ = q^+H^{(2)}, \quad q^-H^{(1)} = H^{(2)}q^-,\label{1}\\
H^{(i)} = -\hbar^2\triangle + V(\vec x),
\quad  \triangle \equiv \partial_1^2 + \partial_2^2,\quad
\partial_i \equiv \partial/\partial x_i,\nonumber\\
q^+ = (q^-)^{\dagger} = \hbar^2 g_{ik}{(\vec x)}\partial_i \partial_k + \hbar
C_i(\vec x,\hbar)\partial_i + B(\vec x,\hbar), \nonumber
\ea
where $\hbar$ is the Plank constant and all coefficient functions
are real.

It means that, up to zero modes of $ q^{\pm}, $ spectra of
$H^{(i)}$ coincide and their eigenfunctions:
\ba
\Psi^{(2)} \sim q^-\Psi^{(1)},\quad  \Psi^{(1)} \sim q^+\Psi^{(2)}.\label{2}
\ea

The intertwining relations (\ref{1}) lead to existence of the symmetry
operators $ R^{(1)}, R^{(2)} $ for the Hamiltonians $H^{(1)}, H^{(2)},$
correspondingly:
\ba
[R^{(i)},H^{(i)}] = 0, \quad R^{(1)} = q^+q^-,\quad R^{(2)} = q^-q^+, \quad
i = 1, 2.\label{3}
\ea

In 1-dim case \cite{10} analogous symmetry operators
$ R^{(i)} $ become polynomials of $ H^{(i)} $ with constant
coefficients. The distinguishing peculiarity of 2-dim case is
existence \cite{9} of nontrivial dynamical symmetry operators
$ R^{(i)} $ which are not reduced to functions of the Hamiltonians
$ H^{(i)}.$

It was shown in \cite{3} that for the unit metrics
$ g_{ik}=\delta_{ik} $ operators $ R^{(i)} $ can be written
as second order differential operators (up to a function of
$ H^{(i)} $) and corresponding quantum systems
allow separation of variables in parabolic, elliptic or polar
coordinate systems. For all other metrics $ g_{ik} $ operators
$ R^{(i)} $ are of fourth order in derivatives.

The intertwining relations (\ref{1}) are equivalent to the following system
of differential equations:
\ba
&&\hbar\partial_i C_k + \hbar\partial_k C_i +
\hbar^2\triangle g_{ik} - (V^{(1)} - V^{(2)})g_{ik} = 0;
\nonumber\\ &&\hbar^2\triangle C_i + 2\hbar\partial_i B +
2\hbar g_{ik}\partial_k V^{(2)} - (V^{(1)} -
V^{(2)})C_i=0;\label{4}\\ &&\hbar^2\triangle B +\hbar^2
g_{ik}\partial_k\partial_i V^{(2)} +\hbar C_i\partial_i V^{(2)}
- (V^{(1)} - V^{(2)}) B = 0.\nonumber
\ea
where the metrics $ g_{ik} $ is a quadratic polynomial in $ x_{1},x_{2} $:
\ba
g_{11}
= a x_2^2 + a_1 x_2 + b_1;
\quad
g_{22} = a x_1^2 +
a_2 x_1 + b_2;\quad g_{12}
=-\frac{1}{2}(2 a x_1 x_2 +  a_1 x_1 +
 a_2 x_2) + b_3. \nonumber
\ea

\vspace{.5cm}
\hspace{3ex}
{\bf I.}\quad For the supercharges with Lorentz metrics
$(g_{ik} = diag(1,-1)):$
\be
q^+ = \hbar^2 (\partial_1^2 - \partial_2^2)
+ \hbar C_k \partial_k + B, \label{5}
\ee
a solution of (\ref{4}) can be reduced \cite{9} to a solution
of the system:
\ba
&&\partial_-(C_- F) =
-\partial_+(C_+ F);\label{6}\\
&&\partial_+^2 F = \partial_-^2 F,\label{7}
\ea
where $ C_{1}\mp C_{2}\equiv C_{\pm}(x_{\pm}) $ depend only on
$ x_{\pm},$ respectively. Eq.(\ref{7}) means that the function
$ F $ can be represented as a sum
$ F=F_{1}(x_{+}+x_{-}) + F_{2}(x_{+}-x_{-}).$
The potentials $ V^{(1,2)} $ and the function $ B $ are expressed
in terms of solutions of system (\ref{6}), (\ref{7}):
\ba
V^{(1,2)}&=&\pm\frac{\hbar}{2}(C_+' + C_-') + \frac{1}{8}(C_+^2 + C_-^2) +
\frac{1}{4}(F_2(x_+ -x_-) - F_1(x_+ + x_-)),\label{8}\\
B&=&\frac{1}{4}(C_+ C_- + F_1(x_+ + x_-) + F_2(x_+ - x_-)).\label{9}
\ea

The solutions for functions $ F, $ which admit additionally the
factorization
$ F=F_{+}(x_{+})\cdot F_{-}(x_{-}),$ were found in \cite{14}.
In the present paper other solutions of (\ref{6}) - (\ref{7})
will be built.

1) After substitution of the general solution of (\ref{6})
\be
F = L\biggl(\int\frac{dx_+}{C_+} - \int\frac{dx_-}{C_-}\biggr)/
(C_+C_-),\label{10}
\ee
into (\ref{7}), we obtain the functional-differential equation
for functions $ L $ and
$ A^{\prime}_{\pm}\equiv 1/C_{\pm}(x_{\pm}):$
\be
\biggl(\frac{A_+'''}{A_+'} - \frac{A_-'''}{A_-'}\biggr)L(A_+ - A_-) +
3 (A_+'' + A_-'')L'(A_+ - A_-) + (A_+'^2 - A_-'^2)L''(A_+ - A_-) = 0,
\label{11}
\ee
where $ L^{\prime} $ denotes the derivative of $ L $ with respect to
its argument.
Eq.(\ref{11}) can be easily solved for functions $ A_{\pm} $ such that
$A_{{\pm}}'' = \lambda^2 A_{\pm},$
Then
\ba
L(A_+ - A_-) = \alpha (A_+ - A_-)^{-2} + \beta,\nonumber
\ea
for $ A_{\pm}=\sigma_{\pm}exp(\lambda x_{\pm}) +
\delta_{\pm}exp(-\lambda x_{\pm})$ with
$ \sigma_{+}\cdot\delta_{+} = \sigma_{-}\cdot\delta_{-}$
and $ \alpha , \beta -$ real constants.
For $ \lambda^{2}>0 $ we obtain (up to an arbitrary shift in $ x_{\pm} $)
two solutions:
\ba
1a)\quad A_{\pm} = k \sinh (\lambda x_{\pm}),\quad
1b)\quad A_{\pm} = k \cosh (\lambda x_{\pm}).\nonumber
\ea
Then (\ref{10}) leads to:
\ba
1a)\quad F_1(2x) = F_2(2x) =\frac{k_1}{\cosh^2(\lambda x)} +
k_2\cosh(2\lambda x),
\quad C_{\pm} = \frac{k}{\cosh(\lambda x_{\pm})},\quad
k\not=0,\label{12}\\
1b)\quad F_1(2x) = - F_2(2x) = \frac{k_1}{\sinh^2(\lambda x)} +
k_2\sinh^2(\lambda x),\quad
C_{\pm} = \frac{k}{\sinh(\lambda x_{\pm})},\quad k\not=0.\label{13}
\ea
For $ \lambda^{2}<0 $ hyperbolic functions must be substituted by
trigonometric ones.

At last, in the limiting case of $ \lambda =0 $ the solutions have the
form:
\ba
&&F_1(2x) = - F_2(2x) = k_1 x^{-2} + k_2 x^2, \quad C_{\pm}
= \frac{k}{x_{\pm}},\quad k\not=0,
\label{14}\\
&&F_1(2x) = - F_2(2x) = k_1 x^2 + k_2 x^4, \quad C_{\pm}
= \pm\frac{k}{x_{\pm}},\quad k\not=0.
\label{15}
\ea

2) To find another class of solutions of the system
(\ref{6}), (\ref{7}) it is useful to replace in (\ref{10})
$ C_{\pm} $ by
$ f_{\pm},$ such that $ C_{\pm}\equiv\pm f_{\pm}/f_{\pm}^{\prime}.$
Then $ F $ in (\ref{10}) is represented in the form
$ F=U(f_{+}f_{-})f_{+}^{\prime}f_{-}^{\prime}$ with an arbitrary
function $ U.$ After substitution in (\ref{7}) one obtains the equation:
\ba
(f_+'^2 f_-^2 - f_+^2 f_-'^2)U''(f) + 3 f \biggl(\frac{f_+''}{f_+} -
\frac{f_-''}{f_-}\biggr)U'(f) +
\biggl(\frac{f_+'''}{f_+'} - \frac{f_-'''}{f_-'}\biggr)U(f) = 0,
\quad f\equiv f_+f_-.\nonumber
\ea

One can check that
$ f_{\pm}=\alpha_{\pm}exp(\lambda x_{\pm}) +
\beta_{\pm}exp(-\lambda x_{\pm}) $ and
$ U=a+4bf_{+}f_{-} $ are its particular solutions
$( a,b -$real constants). Then functions
\ba
&&F_1(x) = k_1(\alpha_+\alpha_-\exp(\lambda x) +
\beta_+\beta_-\exp(-\lambda x))
+ k_2(\alpha_+^2\alpha_-^2\exp(2\lambda x) +
\beta_+^2\beta_-^2\exp(-2\lambda x)),\nonumber\\
&&-F_2(x) = k_1(\alpha_+\beta_-\exp(\lambda x) +
\beta_+\alpha_-\exp(-\lambda x)) +
 k_2(\alpha_+^2\beta_-^2\exp(2\lambda x) +
\beta_+^2\alpha_-^2\exp(-2\lambda x)),\nonumber\\
&&C_{\pm} = \pm \frac{\alpha_{\pm}\exp(\lambda x_{\pm}) +
\beta_{\pm}\exp(-\lambda x_{\pm})}
{\lambda(\alpha_{\pm}\exp(\lambda x_{\pm}) -
\beta_{\pm}\exp(-\lambda x_{\pm}))} \label{16}
\ea
are real solutions of the system (\ref{6}), (\ref{7}) if
$ \alpha_{\pm}, \beta_{\pm} $ are real for the case
$ \lambda^{2}>0 $ and $ \alpha_{\pm}=\beta_{\pm}^{*} $ for the case
$ \lambda^{2}<0.$

3) To find a third class of solutions it is useful to rewrite (\ref{6})
in terms of the variables $ x_{1,2}:$
\ba
2(F_1(x_1) + F_2(x_2))\partial_1(C_+ + C_-) + F_1'(x_1)(C_+ + C_-) +
F_2'(x_2)(C_+ - C_-) = 0.\nonumber
\ea
Its solutions are:
\ba
3a)&&C_+(x_+) = \sigma_1\sigma_2\exp(\lambda x_+) +
\delta_1\delta_2\exp(-\lambda x_+) + c,\nonumber\\
&&C_-(x_-) = \sigma_1\delta_2\exp(\lambda x_-) +
\sigma_2\delta_1\exp(-\lambda x_-) + c,\nonumber\\
&&F_1(x_1) = 0,\quad
F_2(x_2) = \frac{1}{(\sigma_2\exp(\lambda x_2) -
\delta_2\exp(-\lambda x_2))^2};\label{17}\\
3b)&&C_+(x) = C_-(x) = a x^2 + c,\quad F_1(x_1) = 0,\quad
F_2(x_2) = \frac{4b^2}{x_2^2};\label{18}\\
3c)&&C_+(x_+) = \sigma_1\sigma_2\exp(\lambda x_+) +
\delta_1\delta_2\exp(-\lambda x_+),\nonumber\\
&&C_-(x_-) = \sigma_1\delta_2\exp(\lambda x_-) +
\sigma_2\delta_1\exp(-\lambda x_-),\nonumber\\
&&F_{1,2}(x_{1,2}) = \frac{\nu_{1,2}}{(\sigma_{1,2}\exp(\lambda
x_{1,2}) \pm
\delta_{1,2}\exp(-\lambda x_{1,2}))^2} \pm \gamma.
\label{19}
\ea
Let us remark that two additional solutions, analogous to 3a) and 3b),
can be obtained by replacing $ F_{1}(x_{1}) $ with $ F_{2}(x_{2}) $
and vice versa.

After inserting these solutions (\ref{12}) - (\ref{19}) into the general
formulae for potentials (\ref{8}), we obtain, correspondingly,
the following expressions for potentials (\ref{20})-(\ref{27}):
\ba
V^{(1,2)}&=&\mp\frac{\hbar k \lambda}{2}\biggl[\frac{\sinh(\lambda x_+)}
{\cosh^2(\lambda x_+)} + \frac{\sinh(\lambda x_-)}
{\cosh^2(\lambda x_-)}\biggr] + \frac{k^2}{8}\biggl[\frac{1}{\cosh^2(\lambda x_+)} +
\frac{1}{\cosh^2(\lambda x_-)}\biggr] +\nonumber\\&&
 \frac{1}{4}\biggl[\frac{k_1}{\cosh^2(\lambda x_2)} -
\frac{k_1}{\cosh^2(\lambda x_1)} + k_2\cosh(2\lambda x_2) -
k_2\cosh(2\lambda x_1)\biggr];
\label{20}
\ea
\ba
V^{(1,2)}&=&\mp\frac{\hbar k \lambda}{2}\biggl[\frac{\cosh(\lambda x_+)}
{\sinh^2(\lambda x_+)} + \frac{\cosh(\lambda x_-)}
{\sinh^2(\lambda x_-)}\biggr] + \frac{k^2}{8}
\biggl[\frac{1}{\sinh^2(\lambda x_+)} +
\frac{1}{\sinh^2(\lambda x_-)}\biggr] -\nonumber\\&&
 \frac{1}{4}\biggl[\frac{k_1}{\sinh^2(\lambda x_2)} +
\frac{k_1}{\sinh^2(\lambda x_1)} + k_2\cosh(2\lambda x_1) +
k_2\cosh(2\lambda x_2)\biggr];
\label{21}
\ea
\ba
V^{(1,2)}&=&\mp\frac{\hbar k}{2}\biggl(\frac{1}{x_+^2} + \frac{1}{x_-^2}
\biggr) +
\frac{k^2}{8}\biggl(\frac{1}{x_+^2} + \frac{1}{x_-^2}\biggr) -
\frac{1}{4}\biggl[\frac{k_1}{x_1^2} + \frac{k_1}{x_2^2} + k_2(x_1^2 + x_2^2)
\biggr].\label{22}
\ea
Let us note that the potential (\ref{21}) with
$ k_{2}=0 $ and the potential (\ref{22}) were investigated in the
literature (c.f. for example \cite{11}).
\ba
&&V^{(1,2)}=\mp\frac{\hbar k}{2}\biggl(\frac{1}{x_+^2} - \frac{1}{x_-^2}\biggr)
+ \frac{k^2}{8}\biggl(\frac{1}{x_+^2} + \frac{1}{x_-^2}\biggr) -
\frac{1}{4}\biggl[k_1 (x_1^2 + x_2^2) + k_2(x_1^4 + x_2^4)\biggr];
\label{23}\\
&&V^{(1,2)}=
\frac{2\alpha_+\beta_+ (1 \mp 8\hbar\lambda^{2}) +
\alpha_{+}^{2}\exp(2\lambda x_{+}) + \beta_{+}^{2}\exp(-2\lambda x_{+})}
{8\lambda^{2}(\alpha_{+}\exp(\lambda x_{+}) -
\beta_{+}\exp(-\lambda x_+))^{2}} +\nonumber\\&&
\frac{2\alpha_-\beta_- (1 \pm 8\hbar\lambda^{2}) +
\alpha_{-}^{2}\exp(2\lambda x_{-}) + \beta_{-}^{2}\exp(-2\lambda x_{-})}
{8\lambda^{2}(\alpha_{-}\exp(\lambda x_{-}) -
\beta_{-}\exp(-\lambda x_-))^{2}} -\nonumber\\&&
\frac{1}{4}\biggl[k_{1}(\alpha_{+}\beta_{-}\exp(2\lambda x_{2}) +
\alpha_{-}\beta_{+}\exp(-2\lambda x_{2})) +
k_{2}(\alpha_{+}^{2}\beta_{-}^{2}\exp(4\lambda x_{2}) + \nonumber\\&&
\alpha_{-}^{2}\beta_{+}^{2}\exp(-4\lambda x_{2})) +
k_{1}(\alpha_{+}\alpha_{-}\exp(2\lambda x_{1}) +
\beta_{+}\beta_{-}\exp(-2\lambda x_{1})) + \nonumber\\&&
k_{2}(\alpha_{+}^{2}\alpha_{-}^{2}\exp(4\lambda x_{1}) +
\beta_{+}^{2}\beta_{-}^{2}\exp(-4\lambda x_{1}))\biggr];\label{24}\\
&&V^{(1,2)}=\pm\frac{\hbar\lambda}{2}(\sigma_{1}\exp(\lambda x_{1}) -
\delta_{1}\exp(-\lambda x_{1}))(\sigma_{2}\exp(\lambda x_{2}) +
\delta_{2}\exp(-\lambda x_{2})) +\nonumber\\&&
\frac{1}{8}\biggl[(\sigma_{1}^{2}\exp(\lambda x_{1}) +
\delta_{1}^{2}\exp(-\lambda x_{1}))
(\sigma_{2}^{2}\exp(\lambda x_{2}) +
\delta_{2}^{2}\exp(-\lambda x_{2})) +\nonumber\\&&
2c(\sigma_{1}\exp(\lambda x_{1}) -
\delta_{1}\exp(-\lambda x_{1}))
(\sigma_{2}\exp(\lambda x_{2}) -
\delta_{2}\exp(-\lambda x_{2}))\biggr] + \nonumber\\&&
\frac{1}{4(\sigma_{2}\exp(2\lambda x_{2}) -
\delta_{2}\exp(-2\lambda x_{2}))^{2}};\label{25}\\
&&V^{(1,2)}=\pm2\hbar ax_{1} + \frac{1}{4}\bigl[a^{2}(x_1^{4} + x_{2}^{4} +
6x_{1}^{2}x_{2}^{2}) + ac(x_{1}^{2} + x_{2}^{2})\bigr] +
\frac{b^{2}}{x_{2}^{2}};\label{26}\\
&&V^{(1,2)}=\pm\frac{\hbar\lambda}{2}(\sigma_{1}\exp(\lambda x_{1}) -
\delta_{1}\exp(-\lambda x_{1}))(\sigma_{2}\exp(\lambda x_{2}) +
\delta_{2}\exp(-\lambda x_{2})) + \nonumber\\&&
\frac{1}{8}(\sigma_{1}^{2}\exp(\lambda x_{1}) +
\delta_{1}^{2}\exp(-\lambda x_{1}))
(\sigma_{2}^{2}\exp(\lambda x_{2}) +
\delta_{2}^{2}\exp(-\lambda x_{2})) + \nonumber\\&&
\frac{\nu_{2}}{(\sigma_{2}\exp(2\lambda x_{2}) -
\delta_{2}\exp(-2\lambda x_{2}))^{2}} -
\frac{\nu_{1}}{(\sigma_{1}\exp(2\lambda x_{1}) +
\delta_{1}\exp(-2\lambda x_{1}))^{2}}.\label{27}
\ea

The Hamiltonians with potentials (\ref{20}) - (\ref{27}) possess the
symmetry
operators $ R^{(1)}=q^{+}q^{-}, R^{(2)}=q^{-}q^{+},$ where
$ q^{\pm} $ can be obtained inserting solutions (\ref{12}) - (\ref{19})
into (\ref{5}).

It is necessary to note that there are some singular points of
potentials (\ref{21})-(\ref{27}) on the plane $ (x_{1},x_{2}).$
Therefore both the asymptotics of corresponding wave functions
and their behaviour under the action of supertransformation
operators $ q^{\pm} $ (\ref{5}) and of symmetry operators
$ R^{(1)}, R^{(2)}$ have to be investigated.
In particular, for the potentials (\ref{23}) - (\ref{27})
the operators $ q^{\pm} $ the preserve asymptotics of wave functions in
the singular points, so the operators
$ R^{(1)}, R^{(2)} $ are physical symmetry operators for these
systems. For the potentials (\ref{21}), (\ref{22}) symmetry properties
were discussed before (see for example, \cite{11}).

\hspace{3ex}
{\bf II.}\quad For the supercharges with
degenerate metrics $ g_{ik}=diag(1,0):$
\be
q^+ = \hbar^2\partial_1^2
+ \hbar C_k\partial_k + B \label{28}
\ee
Eqs.(\ref{4}) lead to:
\ba
&&C_1(\vec x) = -x_2 F_1'(x_{1})
+ G_1(x_{1}); \qquad C_2(\vec x) = F_1(x_{1});\nonumber\\ &&V^{(1)} = \hbar
(2G_1' - x_2 F_1'') + \frac{1}{4}x_2^2 (F_1^2)'' - x_2 (F_1G_1)' +
K_1(x_1) + K_2(x_2);\nonumber\\ &&V^{(2)} =
\hbar x_2 F_1'' + \frac{1}{4}x_2^2 (F_1^2)'' - x_2 (F_1G_1)' +
K_1(x_1) + K_2(x_2);\nonumber\\ &&B = -\frac{\hbar}{2}(G_1' +
x_2 F_1'') + \frac{1}{2}G_1^2 - \frac{1}{2}x_2^2F_1F_1'' + x_2
F_1G_1' - K_1(x_1),\label{29}
\ea
where  the real functions $ F_{1}(x_{1}), G_{1}(x_{1}), K_{1}(x_{1}) $
are solutions of the following system:
\ba
&&-\frac{\hbar^2 G_1'''}{2} + \frac{\hbar}{2}((G_1^2)'' +
2G_1'^2) + G_1K_1' + 2G_1'K_1 - F_1(F_1G_1)' -
G_1'G_1^2 = m_1F_1;\label{30}\\
&&\frac{\hbar^2}{2}F_1^{(IV)} -
\hbar(F_1'G_1'' + 2G_1'F_1'') - G_1(2G_1'F_1' + G_1''F_1)
-\nonumber\\&&F_1'K_1' + \frac{1}{2} F_1(F_1^2)'' - 2G_1'^2F_1 -
2F_1''K_1 = m_2F_1;\label{31}\\
&&\frac{1}{4}G_1(F_1^2)''' + F_1'(F_1G_1)'' + 3G_1'F_1F_1'' =
m_3F_1;\label{32}\\
&&\frac{1}{4}F_1'(F_1^2)''' + F_1F_1''^2 = m_4F_1, \label{33}
\ea
and $ K_{2}(x_{2}) $ is the polynomial of $ x_{2} $ with constant
coefficients:
\ba
K_2(x_2) = m_0  - m_1 x_2 - \frac{1}{2}m_2 x_2^2 - \frac{1}{3}m_3 x_2^3
+ \frac{1}{4}m_4 x_2^4.\nonumber
\ea

Several particular solutions of Eq.(\ref{33}) can be found. The constant
function $F_{1}=k_{1}$ is the solution of (\ref{33}) for $ m_{4}=0.$
To find other solutions
we define the new function $ U(F_{1}): $
\be
U(F_1) = F_1'(x_1),\label{34}
\ee
to decrease the order of the differential eq. (\ref{33}):
\be
U'' + \frac{3}{U}U'^2
+ \frac{3}{F_1}U' - \frac{2m_4}{U^3} = 0. \label{35}
\ee
Inserting its known solution \cite{17} into (\ref{34}),
the following equation for $ F_{1}(x_{1})$ is obtained:
\ba
\int F_1^{1/2}(m_4 F_1^4 + n F_1^2 + k)^{-1/4} d
F_1 = x_1, \quad n,k=Const. \label{36}
\ea

The integral (\ref{36}) can be written as a finite combination of
elementary functions only in the case when two of three constants
$ m_{4},n,k $ are zero. Thus the solutions of Eq.(\ref{33}) in elementary
functions are: $ F_{1}=k_{1}; F_{1}=x_{1}/n;
F_{1}=(3/2)^{2/3}k^{1/6}x_{1}^{2/3}; F_{1}=m_{4}^{1/2}x_{1}^{2}/4.$
Below, for simplicity, we shall consider the solutions with
particular values of constants $ m_{4},n,k,$ while solutions
with arbitrary values of these constants will differ by some of
coefficients only. To solve Eqs.(\ref{30}) - (\ref{32}) it is
useful to consider separately two cases:
$ G_{1}\equiv 0 $ and $ G_{1}\not\equiv 0.$ In both cases solutions with
$ F_{1}=x_{1} $ lead to potentials $ V^{(1,2)} $ with
separation of variables. Below on such solutions will be ignored.

1) \qquad $ G_{1}=0. $

In this case the potentials $ V^{(1,2)}$ for $ F_{1}=k_{1} $
correspond again to Schr\"odinger equations with separation of variables.
More interesting choices $ F_{1}=x_{1}^{2} $ and $ F_{1}=x_{1}^{2/3}$
lead, respectively, to potentials ($l -$ arbitrary real constant):
\ba
&&V^{(1,2)}=\mp 2\hbar x_2 + lx_{1}^{-2} +
\frac{1}{2}(x_{1}^{4} + 6x_1^2x_2^2 +
8x_{2}^{4}) - \frac{m_2}{8}(x_1^2 + 4x_2^2); \label{37}\\
&&V^{(1,2)}=\frac{7\hbar^2}{36}x_1^{-2} \pm
\frac{2\hbar}{9}x_2x_1^{-4/3} + \frac{1}{9}x_1^{-2/3}(x_{2}^{2} +
\frac{9}{2}x_{1}^{2})
9lx_1^{2/3} - \frac{m_2}{8}(9x_1^2 + 4x_2^2).\label{38}
\ea

2) \qquad $ G_{1}\neq0. $

2a) For $ F_{1}=k_{1}\neq 0 $ Eq.(\ref{31}) leads to the following
equation for $G_{1}(x_{1}):$
\ba
\int\frac{G_1^2 dG_1}{\sqrt{k - \frac{1}{2}m_2G_1^4}} = x_1,
\nonumber
\ea
which has the solutions in terms of elementary functions in two
cases: when $ k>0, m_{2}=0 $ or $ k=0, m_{2}<0. $ For the first one,
after redefinition of constants and translation in $ x_{2},$
\ba
V^{(1,2)}&&= -\frac{5\hbar^2}{36}x_1^{-2} +
\frac{\hbar k_2}{3}(1 \pm 1)x_1^{-2/3} - \frac{k_1k_2}{3}x_2x_1^{-2/3}
+\nonumber\\
&&\frac{1}{4}(k_2^2 +
\frac{3k_1m_1}{k_2})x_1^{2/3} - m_1x_2 +
\frac{k_1^2}{2}. \label{39}
\ea
The second case $ ( k=0, m_{2}<0 ) $ leads to separation of variables.

2b) If
$ F_{1}=x_{1}^{2/3} $ and $ F_{1}=x_{1}^{2} $ the general solutions
of Eq.(\ref{32}) can be found and after substitution into (\ref{30}),
(\ref{31}) give the function $ K_{1}(x_{1}).$ Corresponding
potentials are:
\ba
V^{(1,2)}&=&\frac{7\hbar^2}{36}x_1^{-2} \pm
\frac{2\hbar}{9}x_2x_1^{-4/3} \pm \hbar k_1
+ \frac{1}{9}x_1^{-2/3}(x_{2}^{2} +\frac{9}{2}x_{1}^{2})
-\frac{5k_1}{3}x_2x_1^{2/3} + \nonumber\\&&\frac{3m_1}{8k_1}x_1^{2/3} +
 \frac{k_1^2}{4}x_1^2 -
m_1x_2 + \frac{k_1}{9}(1 + 15k_1)x_2^2,\label{40}\\
V^{(1,2)}&=&(\frac{3\hbar^2}{4} \mp \hbar k_1 +\frac{k_{1}^{2}}{4})
x_{1}^{-2}
\mp 2\hbar x_2 + \frac{1}{2}(x_{1}^{4} + 6x_2^2 x_1^2 + 8x_{2}^{4}) -
 \nonumber\\
&&\frac{1}{8}(12k^2
+ m_2)(x_1^2 + 4x_2^2) -
\frac{km_2 + 6k^3 + 4 k_1}{4}x_2.\label{41}
\ea

Because all potentials (\ref{37}) - (\ref{41}) are singular at
the $ x_{2} $ axis, similar to the case of Lorentz metrics,
it is necessary to investigate separately the behaviour of wave
functions at $ x_{1}\to 0.$
Straightforward though cumbersome calculations show that the
fourth order
differential operators $ R^{(1)}, R^{(2)} $ (see (\ref{3}))
preserve the asymptotics of wave functions for systems (\ref{37}) --
(\ref{41}) and play the role of true dynamical symmetry operators.

\vspace*{0.5cm}
\section{\bf Construction of 2-dim integrable classical systems by
the limit $ \hbar\to 0 $}

Quantum dynamical symmetries which were found by the intertwining method
in Sect.2 have their natural analogs - integrals of motion - in the
corresponding classical systems. These integrals of motion are
polynomials of fourth order in momenta. Similar to Sect.2, it is useful
to consider separately the classical limits for Lorentz and degenerate
metrics.

{\bf I.}\quad For the Lorentz metrics in the limit
$ \hbar\to 0 $ classical supercharge functions become:
\ba
q^{\pm}_{cl} = - 4p_+p_- \pm i(C_-(x_-) p_+ + C_+(x_+) p_-) + B(x_-,x_+).
\nonumber
\ea From Eqs.(\ref{8}) and (\ref{3}) we find that the classical Hamiltonian
\ba
h_{cl} = 2(p_+^2 + p_-^2) + \frac{1}{8}(C_+^2 + C_-^2) +
\frac{1}{4}[F_2(x_+ - x_-) - F_1(x_+ + x_-)],\label{42}
\ea
has the additional integral of motion:
\be
I = 16p_+^2p_-^2 + C_+^2p_-^2 + C_-^2p_+^2 - 2(F_1 + F_2)p_+p_-
+ B^2.\label{43}
\ee

Such sort of classical systems was considered in the literature
(see \cite{13} and references therein). Usually
the functional equation, which
provides existence of integrals of motion for corresponding
classical systems, is solved by the Lax method.
Comparison of (\ref{42}) and (\ref{43}) with notations in \cite{13}
leads to the relations:
\ba
v_1 \equiv 1/8 C_+^2(x_1), v_2 \equiv 1/8 C_-^2(x_2),
v_3(x_{-}) \equiv 1/16 F_2(x_-),
v_4(x_{+}) \equiv - 1/16 F_1(x_+), \label{44}
\ea
and functions $ v_{k} $ must satisfy \cite{13} a certain functional
equation (see Eq.(\ref{47}) below). In the next Section we prove
that (\ref{44}), where $ C_{\pm}, F_{1,2} $ are solutions of
the system (\ref{6}), (\ref{7}), satisfy also Eq.(\ref{47}).
Moreover
some additional solutions of this equation will be found in Sect.4.

{\bf II.}\quad Let us study the integrable classical systems which
can be obtained in the limit $ \hbar\to 0 $ from SSQM systems
with degenerate metrics in $ q^{\pm}.$ The classical supercharges
have the form $ q^{+}_{cl}=(q^{-}_{cl})^{*}=-p_{1}^{2}+
iC_{k}(\vec{x})p_{k}+B(\vec{x}) $ and the Hamiltonian
\ba
H_{cl} = p_k^2 + \frac{1}{2}x_2^2\partial_1^2F_1^2
- x_2(F_1G_1)' + K_1(x_1) + K_2(x_2)\label{45}
\ea
has the integral of motion of fourth order in momenta:
\ba
I \equiv q_{cl}^+q_{cl}^- = p_1^4 + ( C_1^2 - 2B ) p_1^2 + C_2^2
p_2^2 + 2C_1C_2 p_1p_2 + B^2.\label{46}
\ea
All functions in (\ref{45}), (\ref{46}) were defined in the previous
Section, where we have to put $ \hbar =0.$

Thus in the case of degenerate metrics the following classical
integrable systems are obtained (new definitions of constants were
used for some of these systems):
\ba
1)&&V=\frac{1}{2}(x_1^4 + 6x_1^2x_2^2 + 8x_2^4) +
m(x_1^2 + 4x_2^2) + lx_1^{-2},\nonumber\\
&&I= p_1^4 + (6x_1^2x_2^2 + mx_1^2 + x_1^4 + 2lx_1^{-2})p_1^2 +
x_1^4p_2^2 - \nonumber\\&&
4x_1^3x_2p_1p_2 + (x_1^2x_2^2 + mx_1^2 + \frac{x_1^4}{2} +
lx_1^{-2})^{2};\nonumber
\ea
\ba
2)&&V=\frac{1}{9}x_1^{-2/3}(\frac{9}{2}x_1^2 + x_2^2) +
m(9x_1^2 + 4x_2^2) + k x_1^{2/3},\nonumber\\
&&I=p_1^4 + (\frac{2}{9}x_2^2x_1^{-2/3} + 2kx_1^{2/3} + 18mx_1^2
+ x_1^{4/3})p_1^2 + x_1^{4/3}p_2^2 - \nonumber\\
&&\frac{4}{3}x_2x_1^{1/3}p_1p_2 +
(\frac{1}{9}x_2^2x_1^{-2/3} - kx_1^{2/3} -9mx_1^2 -
\frac{1}{2}x_1^{4/3})^2;\nonumber
\ea
\ba
3)&&V=-\frac{k_1k_2}{3}x_2x_1^{-2/3} +
\frac{1}{4}(k_2^2 + \frac{3k_1m_1}{k_2})x_1^{2/3} - m_1x_2,\nonumber\\
&&I=p_1^4 + [\frac{1}{2}(k_2^2 + \frac{3k_1m_1}{k_2})x_1^{2/3}
- \frac{2k_1k_2}{3}x_2x_1^{-2/3} + k_1^2]p_1^2 + k_1^2p_2^2 +
\nonumber\\
&&2k_1k_2x_1^{1/3}p_1p_2 + [\frac{k_1k_2}{3}x_2x_1^{-2/3} +
\frac{1}{4}(k_2^2 - \frac{3k_1m_1}{k_2})x_1^{2/3} - \frac{k_1^2}{2}]^2;
\nonumber
\ea
\ba
4)&&V=\frac{1}{9}x_1^{-2/3}(\frac{9}{2}x_1^2 + x_2^2) -
\frac{5k}{3}x_2x_1^{2/3} + \frac{3m}{8k}x_1^{2/3} +
\frac{k^2}{4}x_1^2 - m x_2 + \frac{k}{9}(1 + 15k)x_2^2,\nonumber\\
&&I=p_1^4 + (\frac{2}{9}x_2^2x_1^{-2/3} - \frac{10k}{3}x_2x_1^{2/3}
+ \frac{k^2}{2}x_1^2 + \frac{3m_1}{4k}x_1^{2/3} +
x_1^{4/3})p_1^2
+ x_1^{4/3}p_2^2 +\nonumber\\ &&2x_1^{2/3}(-\frac{2}{3}x_2x_1^{-1/3} +
k x_1)p_1p_2 +
(\frac{1}{9}x_2^2x_1^{-2/3} + k x_2x_1^{2/3} -
\frac{3m_1}{8k}x_1^{2/3} - \frac{1}{2}x_1^{4/3} +
\frac{k^2}{4}x_1^2)^2;\nonumber
\ea
\ba
5)&&V=\frac{1}{2}(x_1^4 + 6x_1^2x_2^2 + 8x_2^4) +
m(x_1^2 + 4x_2^2) +
\frac{k_1^2}{4}x_1^{-2} -(k_1 -2mk - \frac{3}{2}k^3)x_2,
\nonumber\\
&&I=p_1^4 + \bigl(6x_1^2x_2^2 + 2(m + k^2)x_1^2 -
2k_1x_2 - 4kx_1^2x_2 + x_{1}^{4}
+\frac{k_1^2}{2}x_1^{-2} + kk_1\bigr)p_1^2 +\nonumber\\&&x_1^4p_2^2 +
2x_1(k_1 - 2x_1^2x_2)p_1p_2 +
\bigl(x_1^2x_2^2 + (m + \frac{k^2}{2})x_1^2 +
\frac{1}{2}x_1^4 + k_1x_2 -
\frac{k_1^2}{4}x_1^{-2} - \frac{kk_1}{2}\bigr)^2.
\nonumber
\ea
These potentials are not new: they were found by other methods
in \cite{14} - \cite{16}.

In conclusion of this Section we formulate the procedure of
construction of integrals of motion in terms of Classical Mechanics
objects. For any classical Hamiltonians $ H_{cl} $ and a complex function
$ q^{+}_{cl}(\vec{x},\vec{p})=(q_{cl}^{-})^{*},$ polynomial
in momenta, such that:
\ba
\{q_{cl}^+, H_{cl}\} = i f(\vec x,\vec p)q_{cl}^+,\quad
\{(q_{cl}^+)^*, H_{cl}\} = - i f(\vec x,\vec p)(q_{cl}^+)^*;\nonumber
\ea
with arbitrary real function $ f(\vec{x},\vec{p}),$
 the classical factorizable integral of motion
$ I=q^{+}_{cl}\cdot q^{-}_{cl} $ exists
$( \{ , \} -$ Poisson brackets).

\vspace*{0.5cm}
\section{\bf Integrable systems connected with the Lax method.}

Let us consider classical systems with potentials of the form:
\ba
V(x_1,x_2) = v_1(x_1) + v_2(x_2) + v_3(x_1-x_2) +
v_4(x_1+x_2). \nonumber
\ea
It is known \cite{13} that these systems have the
integrals of motion of fourth order in momenta:
\ba
I = \frac{1}{2}p_1^2p_2^2 + v_{2}(x_{2})p_1^2 - \bigl(v_{3}
(x_{1}-x_{2})-v_{4}(x_{1}+x_{2})\bigr)p_1p_2
+ v_{1}(x_{1})p_2^2 + f,\nonumber
\ea
if the functions $ v_{1}, v_{2}, v_{3}, v_{4} $ satisfy the basic
functional equation:
\ba
&&[v_4(x_1 + x_2) - v_3(x_1 - x_2)][v_2''(x_2) - v_1''(x_1)] +
2[v_4''(x_1 + x_2) - v_3''(x_1 - x_2)]\cdot \nonumber
\\&&[v_2(x_2) - v_1(x_1)] +
3v_4'(x_1 + x_2)[v_2'(x_2) - v_1'(x_1)] +\nonumber\\
&&3v_3'(x_1 - x_2)[v_2'(x_2) + v_1'(x_1)] = 0.\label{47}
\ea

There is a list of known particular solutions of Eq.(\ref{47}) in the
book \cite{13}. To search for new solutions of this equation it
is useful to rewrite it in the equivalent form:
\be
\partial_2(v v_2' + 2v_2\partial_2 v) =
\partial_1(v v_1' + 2v_1\partial_1 v),\label{48}
\ee
where
\be
v \equiv v_4(x_1 + x_2) - v_3(x_1 - x_2).\label{49}
\ee
The general solution of Eq.(\ref{48}) is:
\be
v = \biggl[G\biggl(\int\frac{dx_1}{\sqrt v_1} + \int\frac{dx_2}{\sqrt v_2}
\biggr) +
L\biggl(\int\frac{dx_1}{\sqrt v_1} - \int\frac{dx_2}{\sqrt v_2}\biggr)\biggr]
/{\sqrt{v_1v_2}}; \quad v_{1}\not\equiv 0; \quad v_{2}\not\equiv 0,
\label{50}
\ee
where $ G $ and $ L $ are arbitrary functions of their arguments. Thus
the problem is reduced to searching for the functions $ v $ of the
form (\ref{50}), which satisfy the condition
\ba
(\partial_{1}^{2} - \partial_{2}^{2})v = 0.\label{51}
\ea
In particular, it is easy to check that all solutions which were found
from SSQM in $ \hbar\to 0 $ limit in the Sect.3 (see Eq.(\ref{44})) are
particular solutions of Eq.(\ref{47}) and have the form (\ref{50}) with
$ G\equiv 0. $

Let us apply the technique, which was used in investigation of the system
(\ref{6}), (\ref{7}) in the framework of SSQM, to find the new
particular solutions of eq. (\ref{48}).

1) If the function $ v $ is factorizable
$ v=u_{1}(x_{1})\cdot u_{2}(x_{2}), $ Eq.(\ref{48}) admits separation
of variables and its solutions have the form:
\ba
&&v_k = \frac{n_k[a_k\exp(\sqrt{\lambda}x_k) +
b_k\exp(-\sqrt{\lambda}x_k)] +l_k}{(a_k\exp(\sqrt{\lambda}x_k) -
b_k\exp(-\sqrt{\lambda}x_k))^2}, \quad
k = 1, 2.\nonumber\\
&&v_3 = a_1b_2 \exp(\sqrt{\lambda}\cdot x_{-})
+ a_2b_1 \exp(-\sqrt{\lambda}\cdot x_{-}),\,\nonumber\\
&&v_4 = a_1a_2 \exp(\sqrt{\lambda}\cdot x_{+})
+ b_1b_2 \exp(-\sqrt{\lambda}\cdot x_{+}),\label{*}
\ea
where for $ \lambda >0 $ all constants are real and for $ \lambda <0 $
$ a_{k}=b_{k}^{*}$ and $n_{k},l_{k}$ are real.

2) Let us introduce new functions
\ba
v_1\equiv (W_1')^{-2}, \qquad v_2\equiv (W_2')^{-2}.\label{**}
\ea
Then Eq.(\ref{51}) means that
(derivatives of $ G $ and $ L $ are taken in their arguments):
\ba
&&(G + L)\biggl[\frac{W_1'''}{W_1'} - \frac{W_2'''}{W_2'}\biggr] +
3\biggl[(W_1'' - W_2'')G' + (W_1'' + W_2'')L'\biggr] + \nonumber\\&&
(W_1'^2 - W_2'^2)(G'' + L'') = 0. \label{52}
\ea
It is possible to construct several particular solutions of (\ref{52}).

$2a)$
\be
W_{1,2} = \sigma_{1,2}\exp(\lambda x_{1,2}) +
\delta_{1,2}\exp(-\lambda x_{1,2}),\label{53}
\ee
where constant $ \lambda^{2} $ is real and $ \sigma_{k},\delta_{k} $
are complex. If these constants satisfy the condition
\be
\sigma_1\delta_1 = \sigma_2\delta_2,\label{54}
\ee
then substitution of (\ref{53}) into (\ref{52}) leads to an equation with
separable variables. Its solutions are:
\ba
G(W_1 + W_2)=\frac{\alpha_1}{(W_1 + W_2)^2} + \alpha (W_1 + W_2)^2 +
\beta_1;\label{55}\\
L(W_1 - W_2)=\frac{\alpha_2}{(W_1 - W_2)^2} + \alpha (W_1 - W_2)^2 +
\beta_2,\label{56}
\ea
with arbitrary constants $ \alpha,\alpha_{i},\beta_{i} (i=1,2). $

Let us consider the case with $ \lambda^{2}>0. $ Constants
$ \sigma_{1}, \delta_{1} $ must be both real or both positive because
of requirement that functions $ v_{n}(x_{n}) $ for $ n=1,2 $ should
be real.
Analogous arguments work for the pair $ \sigma_{2}, \delta_{2}.$
The condition (\ref{54}) for real
$ (\sigma_{i},\delta_{i)} $ leads to two options:
\ba
&&W_1(x) = W_2(x) = k \cosh(\lambda x),\label{57}\\
&&W_1(x) = W_2(x) = k \sinh(\lambda x),\quad k \in R,\label{58}
\ea
and the function $ v $ is real for real constants
$ \alpha,\alpha_{i},\beta_{i} (i=1,2). $

In the case of real $ (\sigma_{1},\delta_{1)} $
and imaginary $ (\sigma_{2},\delta_{2)} $ the solutions take the form:
\ba
&&W_1 = k \sinh(\lambda x_1),
\quad W_2 = i k \cosh(\lambda x_2),\label{59}\\
&&W_1 = k \cosh(\lambda x_1),\quad W_2 = i k \sinh(\lambda x_2),
\quad k \in R.\label{60}
\ea
Let us remark that in this case the arguments of the functions
$ G $ and $ L $ are complex conjugated and in order to have a real
function $ v $ the following relation must be fulfilled:
\ba
L^*(W_1 + W_2) = - G(W_1 + W_2).\nonumber
\ea
Therefore
$\alpha_2 = - \alpha_1^*;\quad \beta_2 = - \beta_1^*;\quad \alpha =
\alpha^*$ in (\ref{55}), (\ref{56}).

$ 2b)$
\be
\qquad W_1 = W _2 = g x^2;\quad g^2 \in R.\label{61}
\ee
It follows from Eq.(\ref{52}) that:
\ba
G = a_0 + a_1 (W_1 + W_2) + a (W_1 + W_2)^2;\quad
L = b_0 - \frac{a}{4}(W_1 - W_2)^2 + \frac{b}{(W_1 + W_2)^2},\nonumber
\ea
where all constants are real.

Thus the functions $ v_{n}\quad (n=1,..,4) $ for solutions (\ref{57}) --
(\ref{61}) are, correspondingly:
\ba
&&v_1(x) = v_2(x) = \frac{k}{\sinh^2(\lambda x)},\nonumber\\
&&v_3(x) = v_4(x) = \frac{k_1}{\sinh^2(\lambda x)} +
\frac{k_2}{\sinh^2(\frac{\lambda x}{2})} + k_3\cosh(2\lambda x) +
k_4\cosh(\lambda x);\label{62}
\ea
\ba
&&v_1(x) = v_2(x) = \frac{k}{\cosh^2(\lambda x)},\nonumber\\
&&v_3(x) = \frac{k_1}{\sinh^2(\lambda x)} +
\frac{k_2}{\sinh^2(\frac{\lambda x}{2})} + k_3\cosh(2\lambda x) +
k_4\cosh(\lambda x);\nonumber\\
&&v_4(x) = \frac{k_1 + 4k_2}{\sinh^2(\lambda x)} -
\frac{k_2}{\sinh^2(\frac{\lambda x}{2})} + k_3\cosh(2\lambda x) -
k_4\cosh(\lambda x);\label{63}
\ea
\ba
&&v_1(x) = \frac{k}{\cosh^2(\lambda x)},\quad
v_2(x) = - \frac{k}{\sinh^2(\lambda x)};\nonumber\\
&&v_3(x) = v_4(x) = \frac{k_1 + k_2\sinh(\lambda x)}{\cosh^2(\lambda x)} +
k_3\cosh(2\lambda x) - k_4\sinh(\lambda x);\label{64}
\ea
\ba
&&v_1(x) = \frac{k}{\sinh^2(\lambda x)},\quad
v_2(x) = - \frac{k}{\cosh^2(\lambda x)};\nonumber\\
&&v_3(x) = \frac{k_1 + k_2\sinh(\lambda x)}{\cosh^2(\lambda x)} +
k_3\cosh(2\lambda x) + k_4\sinh(\lambda x);\nonumber\\
&&v_4(x) = \frac{k_1 - k_2\sinh(\lambda x)}{\cosh^2(\lambda x)} +
k_3\cosh(2\lambda x) - k_4\sinh(\lambda x);\label{65}
\ea
\ba
v_{1}(x) = v_{2}(x) = k x^{-2};\quad
v_{3}(x) = v_{4}(x) = k_{1}x^{-2} + k_{2}x^{2} + k_{3}x^{4} + k_{4}x^{6},
\label{66}
\ea
\ba
&&v_{1} = \frac{g_{1}}{(\delta_{1}\exp(\lambda x) -
\sigma_{1}\exp(-\lambda x))^{2}},\quad
v_{2} = \frac{g_{2}}{(\delta_{2}\exp(\lambda x) -
\sigma_{2}\exp(-\lambda x))^{2}};\label{67}\\
&&v_{3}(x) = k_{1}(\delta_{1}\sigma_{2}\exp(\lambda x) +
\sigma_{1}\delta_{2}\exp(-\lambda x)) +
k_{2}(\delta_{1}^{2}\sigma_{2}^{2}\exp(2\lambda x) +
\sigma_{1}^{2}\delta_{2}^{2}\exp(-2\lambda x));\nonumber\\&&
v_{4}(x) = k_{1}(\delta_{1}\delta_{2}\exp(\lambda x) +
\sigma_{1}\sigma_{2}\exp(-\lambda x)) +
k_{2}(\sigma_{1}^{2}\sigma_{2}^{2}\exp(2\lambda x) +
\delta_{1}^{2}\delta_{2}^{2}\exp(-2\lambda x)).\nonumber
\ea

The solutions (\ref{63}) -- (\ref{65})
are absent in the list of \cite{13} and to our knowledge
they are novel. As to
expressions (\ref{67}), they are present in \cite{13} only for
$ \sigma_{1}\delta_{1}>0,\,\sigma_{2}\delta_{2}>0. $

In conclusion, let us note that one can easily check some invariance
properties of Eq.(\ref{47}). In particular, from arbitrary solutions
(\ref{*}), (\ref{62}) -- (\ref{67}) one derives new ones with:
\ba
v_{4}(2x) \to v_{1}(x); \quad v_{3}(2x) \to v_{2}(x); \quad
v_{1}(x) \to -v_{3}(x); \quad v_{2}(x) \to -v_{4}(x) . \label{x}
\ea
Eq.(\ref{47}) is invariant if
$ v_{1,2} \to v_{1,2} + c $ and $ v_{3,4} \to v_{3,4} + \tilde c $
with arbitrary constants $ c, \tilde c.$ It is also invariant under
dilatation of all arguments $ x_{i} \to \Lambda x_{i}. $

\vspace*{0.5cm}
\section{\bf Integrable systems with potentials, expressed
in elliptic functions.}

In this Section we formulate the method of construction of new
integrable systems from solutions $v_i$ of Eq.(\ref{47}), which were
found in previous Section. If we define new functions:
\be
W_{1}'^{2} = f_{1}(W_{1}),\quad W_{2}'^{2} = f_{2}(W_{2}), \label{1x}
\ee
Eq.(\ref{52}) takes the form:
\ba
&&[G(W_1 + W_2) - L(W_1 - W_2)][f_2''(W_2) - f_1''(W_1)] +\nonumber\\
&&2[G''(W_1 + W_2) - L''(W_1 - W_2)][f_2(W_2) - f_1(W_1)] +\nonumber\\
&&3G'(W_1 + W_2)[f_2'(W_2) - f_1'(W_1)] +
3L'(W_1 - W_2)[f_2'(W_2) + f_1'(W_1)] = 0.\label{2x}
\ea
This equation has the same structure as Eq.(\ref{47}), thus
from (\ref{1x}) we obtain:
\ba
&&W_{1}'^{2} = v_{1}(W_{1});\quad W_{2}'^{2} = v_{2}(W_{2});\nonumber\\&&
G(W_1 + W_2) = v_{4}(W_1 + W_2);\quad L(W_1 - W_2) = -v_{3}(W_1 - W_2),
\label{3x}
\ea
where we can use solutions (\ref{*}), (\ref{62}) - (\ref{67})
for $v_n(x)\quad (n=1,...,4)$ with arguments $x_{1,2}$ replaced,
respectively, by $W_{1,2}.$ The first pair of equations in (\ref{3x})
defines functions $ W_{1}(x_{1}), W_{2}(x_{2}) $ and last two equations
define new functions $ G, L. $ Substituting these set of functions
into (\ref{**}) and (\ref{50}), we find
some new solutions $ V_{n} $ of the same Eq.(\ref{47})
from already known solutions
$ v_{n} $ (see (\ref{*}), (\ref{62}) - (\ref{67})).

Let us consider several examples of this method of reproducing new
solutions.

\quad 1)\, The first attempt to start our procedure from the simplest
solutions (\ref{66}) leads to discouraging result: we obtain the same
solution. But we can firstly transform
$ v_{1,2}\Longleftrightarrow v_{3,4} $ in (\ref{66}), using
the invariance property
(\ref{x}) mentioned at the very end of Sect.4. As follows from (\ref{3x}),
the functions $ W_{1,2}(x_{1,2}) $ are defined from the
equations (we omit
indices $ i=1,2 $):
\ba
W'^{2}(x) = k_{1}W^{-2} + k_{2}W^{2} + k_{3}W^{4} + k_{4}W^{6} + k_{0},
\nonumber
\ea
with constant $ k_{i}.$ It is useful to rewrite them in terms of
functions $ U(x)\equiv \frac{1}{2}W^{2}(x): $
\be
U'^{2}(x) =k_{1} + k_{0}U + k_{2}U^{2} + k_{3}U^{3} + k_{4}U^{4}.\label{5x}
\ee From (\ref{50}) and (\ref{**}) we obtain new solutions:
\ba
&&V_{1} = g_{1}\frac{U(x_{1})}{U'^{2}(x_{1})},\quad
V_{2} = g_{1}\frac{U(x_{2})}{U'^{2}(x_{2})},\label{6x}\\
&&V_{4}(x_{1} + x_{2}) - V_{3}(x_{1} - x_{2}) =
g\frac{U'(x_{1})U'(x_{2})}{(U(x_{1}) - U(x_{2}))^{2}},\label{7x}
\ea
where arbitrary constants $ g_{1}, g $ appear because solutions
$ v_{n} $ are defined from Eq.(\ref{47}) up to constant factors.
We can check directly that for $ U(x), $ which satisfy
Eq.(\ref{5x}), r.h.s. of (\ref{7x}) is the solution of Eq.(\ref{51}).

When r.h.s. polynomial in Eq.(\ref{5x}) has degenerate roots,
$ U(x) $ can be expressed through elementary functions and corresponds
to solutions (\ref{*}), (\ref{62}) -- (\ref{67}), found above.
When all roots are simple, $ U(x) $ can be
given in terms of elliptic functions.

\qquad 1a) \quad $U(x) = \wp(x) + b,$\\
where $ b $ is constant and $ \wp(x) $ is the Weierstrass function with
semiperiods $ \omega_{1}, \omega_{2} $ (their values depend on constants
$k_{i}$). Eqs.(\ref{6x}), (\ref{7x}) lead to new solutions:
\ba
&&V_{1}(x) = V_2(x) = a_{1}\wp(x + \omega_{1}) + a_{2}\wp(x + \omega_{2}) +
a_{3}\wp(x + (\omega_{1} + \omega_{2})/2);\nonumber\\&&
V_{3}(x) = V_4(x) = a\wp(x),\nonumber
\ea
where $ a $ is arbitrary constant and constants
$ a_{k} (k=1,2,3) $ satisfy the following condition:
\ba
\sum_{k}a_{k}^{2} - \sum_{i\not=j}a_{i}a_{j} =1.\nonumber
\ea

\quad 1b)\quad $U(x) = {\rm sn}\, x + b,$
where
$ b $ is constant and $ {\rm sn}\,(x)$ is the Jacobi function with
modulus $ k $ (it
depends again on constants $ k_{i} $). In this case new solutions are:
\ba
&&V_{1}(x) = V_{2}(x) = a \frac{{\rm sn}\, x +b}{{\rm cn}^{2}x
\cdot {\rm dn}^{2}x},\nonumber\\
&&V_{3}(x) = c(\frac{1}{{\rm sn}^{2}(x/2)} - k^{2}
{\rm cn}^{2}(x/2) - k^{2}),\,
V_{4}(x) = c(1 - k^{2})(\frac{1}{{\rm cn}^{2}(x/2)} -
\frac{1}{{\rm sn}^{2}(x/2)}).
\nonumber
\ea

\quad 1c)\quad $U(x) = {\rm dn} x + b.$

Correspondingly, the new solution takes the form:
\ba
V_{1}(x) = V_{2}(x) = a \frac{{\rm dn} x +b}
{{\rm sn}^{2}x\cdot {\rm dn}^{2}x};\,
V_{3}(x) = V_{4}(x) = c(\frac{1}{{\rm sn}^{2}(x/2)} +
(1 - k^{2}){\rm cn}^{2}(x/2)).
\nonumber
\ea

\quad 2)\, The second solution of (\ref{47}), which we can take as the
starting point of proposed procedure, is one of solutions in Eq.(\ref{*}):
\ba
&&v_{1}(x)=v_{2}(x)=a \cos x + c,\qquad a > 0,\quad c > 0,\nonumber\\
&&v_{3}(x)=k_{1}(\sin(x/2))^{-2} + k_{2}(\sin(x/4))^{-2},\quad
v_{4}(x)=k_{3}(\sin(x/2))^{-2} + k_{4}(\sin(x/4))^{-2}.\nonumber
\ea

Then according to Eq.(\ref{3x}), functions $ W(x) $ must be found from
equation:
\ba
W'^{2}(x) = a\cdot \cos W(x) + c,\nonumber
\ea
which solution can be expressed through the Jacobi function with modulus
$k \quad (k^{2}\equiv 2a/(a+c)):$
\be
W(x) = \arccos(1 - 2(k\cdot {\rm sn}\, y)^{-2}), \quad y\equiv
\frac{1}{2}\sqrt{(a+c)}\cdot x.
\label{4.1x}
\ee
Thus the new solution of Eq.(\ref{47}) is:
\ba
&&V_{4}(x_{1} + x_{2}) = a_{1}{\rm sn}^{2}(y_{+}/2) + a_{2}\frac{{\rm cn}^{2}(y_{+}/2)}
{{\rm dn}^{2}(y_{+}/2)} + a_{3}\frac{{\rm dn}^{2}(y_{+}/2)}{{\rm sn}^{2}(y_{+}/2){\rm cn}^{2}(y_{+}/2)}
+ a_{4}\frac{{\rm dn}^{2}(y_{+}/2)}{{\rm cn}^{2}(y_{+}/2)},\nonumber\\
&&V_{3}(x_{1} - x_{2}) = a_{2}{\rm sn}^{2}(y_{-}/2) + a_{1}\frac{{\rm cn}^{2}(y_{-}/2)}
{{\rm dn}^{2}(y_{-}/2)} + a_{3}\frac{{\rm dn}^{2}(y_{-}/2)}{{\rm sn}^{2}(y_{-}/2){\rm cn}^{2}(y_{-}/2)}
+ a_{4}\frac{1}{{\rm sn}^{2}(y_{-}/2)},\nonumber\\
&&V_{1}(x) = V_{2}(x) = \frac{a_{0}}{{\rm cn}^{2}y},\nonumber
\ea
where
$ a_{0}, a_{k} $ are arbitrary constants.

In conclusion, we mention briefly the analogous method of construction
of new integrable systems in Quantum  Mechanics (see Sect.2). In this
case the procedure is based on Eq.(\ref{11}) of Sect.2. Similarly to
(\ref{1x}), we introduce in (\ref{11}) new functions $ M_{\pm}(A_{\pm}):$
\be
A_{+}'^{2}(x_{+}) = M_{+}(A_{+}),\quad
A_{-}'^{2}(x_{-}) = M_{-}(A_{-}).\label{5.1x}
\ee
Then Eq.(\ref{11}) takes the form:
\ba
&&[M_{+}''(A_{+}) - M_{-}''(A_{-})]L(A_{+}-A_{+}) +
3[M_{+}'(A_{+}) + M_{-}'(A_{-})]L'(A_{+}-A_{+}) +\nonumber\\
&&2[M_{+}(A_{+}) - M_{-}(A_{-})]L''(A_{+}-A_{+}) = 0.\nonumber
\ea

The general solution of this equation can be found in \cite{17}
and the corresponding functions $ M_{\pm}(A_{\pm}), L(A_{+}-A_{-}) $
can be used here to find $ A_{\pm}(x)$ from (\ref{5.1x}).
Substitution of these functions
$ A_{\pm}(x)$ into $ M_{\pm}(A_{\pm}), L(A_{+}-A_{-}) $ and Eq.(\ref{10})
leads to new solutions of the system Eqs.(\ref{6}), (\ref{7}) for the
quantum case.\\
{\bf Acknowledgements:}\\
This paper was supported by RFBR grant No. 99-01-00736. 
One of the authors
(M.I.) was partially supported by CIMO (Finland).
He is also grateful to University of Helsinki (Prof.M.Chaichian)
for kind hospitality.

\end{document}